\documentclass{article}
\usepackage{float}
\usepackage{graphicx}      
\usepackage{amsmath}       
\usepackage{amssymb}       
\usepackage{amsthm}        
\usepackage{bm}     
\usepackage[hidelinks]{hyperref}     
\usepackage[nameinlink]{cleveref} 
\usepackage{caption}       
\usepackage{subcaption}    
\usepackage{float}         
\usepackage{booktabs}      
\usepackage{longtable}     
\usepackage{multirow}      
\usepackage{authblk}
\usepackage{array}  
\usepackage[version=4]{mhchem}
\usepackage{geometry}      
\usepackage[
backend=biber,
style=numeric,
sorting=none
]{biblatex}

\usepackage{ragged2e}

\title{Chemical Frequency Combs in Reaction-Diffusion Oscillators}

\author[1]{Krishnesh Krishnakumar Nair}
\author[2]{Madhurendra Mishra}
\author[3]{Zhen Qi}
\author[4]{Adarsh Ganesan}

\affil[1]{Department of Electronics and Communication Engineering\\
National Institute of Technology, Warangal, Telangana, India\\
{\normalfont\texttt{kk24ecb0b30@student.nitw.ac.in}}}

\affil[2]{Department of Physics\\
Sri Guru Tegh Bahadur Khalsa College, University of Delhi, Delhi, India\\
{\normalfont\texttt{madhurendramishra24@gmail.com}}}

\affil[3]{Department of Biomedical Engineering\\
Worcester Polytechnic Institute, Worcester, MA 01609, United States\\
{\normalfont\texttt{zqi@wpi.edu}}}

\affil[4]{Department of Electrical and Electronics Engineering\\
BITS Pilani, Dubai Campus, Dubai, UAE\\
{\normalfont\texttt{adarsh@dubai.bits-pilani.ac.in}}}

\date{}

\begin{document}

\maketitle

\begin{abstract}
Frequency combs, evenly spaced spectral lines locked to one fundamental frequency, are well known in optics and have also been found in phononic, magnonic, ferroelectric, and cosmological systems, but have not yet been studied in oscillating chemical reactions. In this work, we show that reaction-diffusion oscillators can also produce frequency combs. We use the Brusselator model and derive its Hopf bifurcation condition directly from the rate equations. We find that the trimolecular autocatalytic term is the only source of nonlinear harmonic content. Above the Hopf threshold, our simulations of target-wave patterns show a clear fundamental frequency followed by a long, evenly spaced ladder of harmonics, with each harmonic weaker than the one before it. We then vary the reactant concentrations and kinetic parameters one at a time and find that this comb structure holds across a wide range of values. We also test this idea experimentally using the Belousov-Zhabotinsky reaction. Intensity signals recorded at different points in a target pattern show a shared fundamental frequency with several weakening harmonics, matching the simulated pattern closely. Together, these results show that reaction-diffusion chemistry is a new platform for generating frequency combs.
\end{abstract}

\section{Introduction}

Chemical systems far from equilibrium can break spatial and temporal homogeneity through the coupling of diffusion to nonlinear kinetics. Turing showed that differential diffusivity between two reacting species destabilises the uniform state and selects a finite wavelength~\cite{Turing1952}. The Belousov--Zhabotinsky reaction confirmed that chemistry alone supports such self-organisation, producing sustained oscillations, spiral waves, and target patterns~\cite{ZaikinZhabotinsky1970,Field1972}.
Boris Belousov made the first demonstration of this reaction in 1951 in a system comprising of potassium bromate, cerium(IV) sulfate, malonic acid, citric acid and dilute sulfuric acid. The ratio of concentration of the cerium(IV) and cerium(III) ions kept oscillating causing the colour of the solution to swing back and forth between a yellow solution and a colorless solution. This experiment after years of debate got published in 1959 \cite{Belousov1959} and later extended by Anatol Zhabotinsky \cite{zhabotinsky1964periodic}. In 1991, stationary Turing patterns were observed in the chlorite-iodide-malonic acid system, and were rationalised by the Lengyel--Epstein model: gel-binding of the activator supplies the diffusion disparity required for the instability~\cite{LengyelEpstein1991}. Coupling a slow feedback step to a bistable substrate in a continuous stirred-tank reactor (CSTR) provided a systematic route to new oscillators and rapidly enlarged the catalogue of chemical clocks~\cite{DeKepperEpstein1981,EpsteinShowalter1996}. Activator-inhibitor reaction-diffusion ideas were later carried over to developmental biology~\cite{KondoMiura2010,MarciniakFrey2016}, and cross-diffusion was shown to broaden the parameter window for pattern formation in realistic chemical settings~\cite{VanagEpstein2009}. The near-equilibrium picture has since been refined: mass-conserving systems are characterised by moving local equilibria far from global equilibrium~\cite{HalatekFrey2018}, and pattern selection can proceed through saddle-node bifurcations below the Turing threshold~\cite{BraunsFrey2020,WuZhang2024}. Even pairs of identical coupled oscillators exhibit robust symmetry breaking, mixed-mode oscillations, and multi-timescale rhythms~\cite{Epstein2014,AwalEpstein2023,EpsteinAwal2024}, and continuing work on oscillating networks keeps revealing new regimes~\cite{CupicEpstein2021,DalwadiPearce2023}.

Frequency combs represent a universal nonlinear mechanism occuring in disparate physical domains. The concept originated in nonlinear optics, where mode locking in a high-finesse cavity produces a dense set of equally spaced, phase-coherent lines~\cite{Udem2002,Cundiff2003}, with transformative impact on precision metrology, spectroscopy, and optical clocks. The underlying mechanism, nonlinear mode coupling onto a phase-coherent grid, is not specific to optics and transfers readily to other domains.

Phononic frequency combs (PFCs) is the first extension of this celebrated concept beyond optics. Ganesan and coworkers generated PFCs through multimode nonlinear coupling in micromechanical resonators~\cite{Ganesan2017}, and follow-up work mapped additional routes via parametric mixing and resonance interactions~\cite{Ganesan2018a,Ganesan2018b}. PFCs have since been reported across a range of experimental platforms~\cite{Chiout2021,Kesekler2022,anderson2026phononic,deJong2023,Xiao2026}, with recent proposals targeting molecular vibrations~\cite{Lei2024}, twisted bilayer van der Waals materials~\cite{Liu2025}, and engineered solid-state architectures~\cite{Rangwala2026}. Complementary analyses establish the existence conditions and mode-coupling requirements for comb spectra~\cite{Qi2020}.

The frequency comb paradigm has been further extended. Magnonic combs arise from three-magnon scattering and parametric spin-wave interactions in ferromagnetic insulators such as yttrium iron garnet~\cite{Demidov2020}. In ferroelectrics, nonlinear polarisation waves, or ferrons, produce terahertz combs whose spacing and efficiency follow from the static polarisation of the underlying modes~\cite{PolarizationComb2026}. Trivedi et al.\ recently identified cosmological combs as time-periodic attractor solutions of exponential quintessence models, in which phase-locked oscillations modulate the Hubble rate and the structure-growth rate~\cite{Trivedi2026}. Equally spaced, phase-coherent spectra in micromechanical resonators, magnetic insulators, ferroelectrics, and cosmological evolution together point to a strong universality of the underlying nonlinear mechanism.

Oscillatory chemical reactions are a natural but unexplored host for comb generation. They produce periodic signals from feedback-driven kinetics far from equilibrium; the nonlinear terms that sustain the limit cycle also generate harmonics of the fundamental, and intermodal mixing yields sidebands and combination tones whenever several modes are coupled or the system is forced periodically~\cite{Epstein1998,Strogatz2014}. The three ingredients required for comb formation, nonlinear coupling, self-consistent energy redistribution, and phase locking, are therefore already in place. However, so far there has been no direct demonstration of the frequency comb concept in chemical systems.

The present work addresses this gap. We define a \emph{chemical frequency comb} as a discrete, nearly equally spaced family of spectral components in the response of an oscillatory chemical reaction, produced by nonlinear coupling between the fundamental oscillation and its harmonic and sideband modes. We establish the conditions for comb formation in representative reaction networks, link comb structure to the underlying kinetic nonlinearities, and show how periodic forcing and parameter variation tune the spectral response.

\section{Theory}
\label{sec:theory}

The standard models for oscillatory chemical kinetics are the three-state variable Oregonator~\cite{Field1972,FieldNoyes1974} and the two-state variable Brusselator~\cite{Prigogine1968}. We adopt the latter. The Brusselator consists of four steps~\cite{Prigogine1968,Nicolis1977}
\begin{align}
\text{A} &\xrightarrow{k_1} \text{X}, \label{eq:bruss_r1}\\
\text{B} + \text{X} &\xrightarrow{k_2} \text{Y} + \text{D}, \label{eq:bruss_r2}\\
2\,\text{X} + \text{Y} &\xrightarrow{k_3} 3\,\text{X}, \label{eq:bruss_r3}\\
\text{X} &\xrightarrow{k_4} \text{E}, \label{eq:bruss_r4}
\end{align}
with A and B held at fixed concentration by continuous feed, D and E inert products, and X, Y the reactive intermediates. The autocatalytic trimolecular step~\eqref{eq:bruss_r3} provides positive feedback in X, while step~\eqref{eq:bruss_r2} supplies the delayed negative feedback by consuming X and regenerating Y. Allowing X and Y to diffuse with coefficients $D_X$ and $D_Y$ extends the rate equations to~\cite{Turing1952,Nicolis1977}
\begin{align}
\frac{\partial[X]}{\partial t} &= k_1[\mathrm{A}] - k_2[\mathrm{B}][X] + k_3[X]^2[Y] - k_4[X] + D_X\nabla^2[X], \label{eq:bruss_pde_X}\\[4pt]
\frac{\partial[Y]}{\partial t} &= k_2[\mathrm{B}][X] - k_3[X]^2[Y] + D_Y\nabla^2[Y]. \label{eq:bruss_pde_Y}
\end{align}
Diffusion leaves the uniform steady state unchanged. For $D_Y \gg D_X$, fast inhibitor diffusion combined with local autocatalysis in $X$ drives a Turing instability and selects stationary spatial patterns~\cite{Nicolis1977}. The trimolecular term $k_3[X]^2[Y]$ is the only nonlinearity in the model, and enables intermodal mixing that populate higher-order spectral components. Far above the Hopf threshold, the limit cycle becomes strongly anharmonic, with comb-like spectra emerging as the natural signature of this regime. The two-variable structure keeps the analysis tractable making the Brusselator model an interesting test bed for chemical frequency combs.

\subsection{Hopf bifurcation condition for the reaction-diffusion system}
\label{sec:hopf_derivation}

To derive the Hopf bifurcation condition for the Brusselator reaction-diffusion system, we define the variables as
\begin{equation}
a = k_1 \left[ A \right], \qquad b = k_2 \left[ B \right], \qquad c = k_3, \qquad d = k_4.
\end{equation}
The reaction terms become
\begin{align}
f(x,y) &= a - b x + c x^2 y - d x, \label{eq:f(x,y)} \\ 
g(x,y) &= b x - c x^2 y. \label{eq:g(x,y)}
\end{align}

\subsubsection{Steady-state solution}

At the steady state,
\begin{equation}
f(x_0,y_0) = 0, \qquad g(x_0,y_0) = 0.
\end{equation}
Hence, we have
\begin{equation}
b x_0 - c x_0^2 y_0 = 0.
\end{equation}
For the nontrivial solution $x_0 \neq 0$,
\begin{equation}
y_0 = \frac{b}{c x_0}.
\end{equation}
Substituting this into the equation \eqref{eq:f(x,y)} gives
\begin{equation}
a - b x_0 + c x_0^2 \left(\frac{b}{c x_0}\right) - d x_0 = 0,
\end{equation}
so that
\begin{equation}
a - d x_0 = 0.
\end{equation}
Therefore,
\begin{equation}
x_0 = \frac{a}{d} = \frac{k_1 \left[ A \right]}{k_4},
\end{equation}
and
\begin{equation}
y_0 = \frac{b}{c x_0} = \frac{k_2 \left[ B \right] k_4}{k_3 k_1 \left[ A \right]}.
\end{equation}

\subsubsection{Linearization of the reaction equations}

Consider small perturbations around the steady state:
\begin{equation}
x = x_0 + \delta x, \qquad y = y_0 + \delta y.
\end{equation}
The Jacobian matrix of the reaction terms is
\begin{equation}
J =
\begin{pmatrix}
\dfrac{\partial f}{\partial x} & \dfrac{\partial f}{\partial y} \\[8pt]
\dfrac{\partial g}{\partial x} & \dfrac{\partial g}{\partial y}
\end{pmatrix}.
\end{equation}
The individual derivatives are
\begin{align}
\frac{\partial f}{\partial x} &= -b + 2 c x y - d, & \frac{\partial f}{\partial y} &= c x^2, \\
\frac{\partial g}{\partial x} &= b - 2 c x y, & \frac{\partial g}{\partial y} &= -c x^2.
\end{align}
At the steady state,
\begin{equation}
c x_0 y_0 = b.
\end{equation}
Therefore,
\begin{equation}
J_0 =
\begin{pmatrix}
b - d & c x_0^2 \\
-b & -c x_0^2
\end{pmatrix}.
\end{equation}
The trace and determinant are
\begin{align}
\operatorname{tr} J_0 &= b - d - c x_0^2, \\
\det J_0 &= c d x_0^2.
\end{align}
Since $c > 0$, $d > 0$, and $x_0 > 0$,
\begin{equation}
\det J_0 > 0.
\end{equation}

\subsubsection{Hopf bifurcation condition}

The eigenvalues satisfy
\begin{equation}
\lambda^2 - \operatorname{tr} J_0 \,\lambda + \det J_0 = 0.
\end{equation}
A Hopf bifurcation occurs when a complex-conjugate pair of eigenvalues crosses the imaginary axis. For a two-variable system, the conditions are
\begin{equation}
\operatorname{tr} J_0 = 0, \qquad \det J_0 > 0.
\end{equation}
Because the determinant is automatically positive, the critical condition is
\begin{equation}
b - d - c x_0^2 = 0.
\end{equation}
Using $x_0 = a/d$, we obtain
\begin{equation}
b = d + c \left(\frac{a}{d}\right)^2.
\end{equation}
Returning to the original parameters,
\begin{equation}
\label{eq:hopf_condition}
k_2 \left[ B \right] = k_4 + k_3 \left(\frac{k_1 \left[ A \right]}{k_4}\right)^2.
\end{equation}
\subsubsection{Treating $\left[ \mathbf{A} \right]$ as the control parameter}

\justify
Here, we keep \(k_{1}\), \(k_{2}\), \(k_{3}\), \(k_{4}\), and \([B]\) fixed while varying \([A]\).

\justify
Rearranging equation~\eqref{eq:hopf_condition}, we get
\begin{align}
k_{2}\left[ B \right] - k_{4}
&= k_{3}\frac{k_{1}^{2}\left[ A \right]^{2}}{k_{4}^{2}}.
\end{align}
\justify

Therefore,
\begin{align}
\left[ A \right]^{2}
&= \frac{k_{4}^{2}}{k_{1}^{2}}
   \frac{k_{2}\left[ B \right] - k_{4}}{k_{3}}.
\end{align}
\justify

Thus,
\begin{align}
\left[ A \right]_{H}
&= \frac{k_{4}}{k_{1}}
   \sqrt{\frac{k_{2}\left[ B \right] - k_{4}}{k_{3}}}.
\end{align}
\justify

This Hopf bifurcation exists only if
\begin{align}
k_{2}\left[ B \right] &> k_{4}.
\end{align}
\justify

If
\begin{align}
k_{2}\left[ B \right] &\leq k_{4},
\end{align}
\justify
then there is no positive real value of \(\left[ A \right]_{H}\).

\justify
Because
\begin{align}
\operatorname{tr}\left( J_{0} \right)
&= k_{2}\left[ B \right] - k_{4}
   - k_{3}\frac{k_{1}^{2}\left[ A \right]^{2}}{k_{4}^{2}},
\end{align}
\justify
increasing \(\left[ A \right]\) decreases the trace. Therefore,
\begin{align}
\left[ A \right] < \left[ A \right]_{H}
&\quad \Rightarrow \quad \operatorname{tr}\left( J_{0} \right) > 0,
\end{align}
\justify
so the steady state is oscillatory unstable. Meanwhile,
\begin{align}
\left[ A \right] > \left[ A \right]_{H}
&\quad \Rightarrow \quad \operatorname{tr}\left( J_{0} \right) < 0,
\end{align}
\justify
so the steady state is stable.

\justify
Therefore, for \(\left[ A \right]\) as the control parameter,
\begin{align}
\left[ A \right]_{H}
&= \frac{k_{4}}{k_{1}}
   \sqrt{\frac{k_{2}\left[ B \right] - k_{4}}{k_{3}}},
\end{align}
\justify
with instability for
\begin{align}
0 < \left[ A \right] &< \left[ A \right]_{H}.
\end{align}
\justify

\subsubsection{Treating $\left[ \mathbf{B} \right]$ as the control parameter}

\justify
Here, we keep \([A]\), \(k_{1}\), \(k_{2}\), \(k_{3}\), and \(k_{4}\)  fixed while varying \([B]\).

\justify

Rearranging equation~\eqref{eq:hopf_condition}, we get
\begin{equation}
\left[ B \right] = \frac{1}{k_2}\left[k_4 + k_3 \left(\frac{k_1 \left[ A \right]}{k_4}\right)^2\right].
\label{eq:BH_dimensional}
\end{equation}
When $\left[ B \right]$ is used as the control parameter,

\begin{equation}
\left[ B \right]_L = \frac{1}{k_2}\left[k_4 + k_3 \left(\frac{k_1 \left[ A \right]}{k_4}\right)^2\right].
\label{eq:BH_dimensional}
\end{equation}
with instability for
\begin{equation}
\left[ B \right] > \left[ B \right]_L
\end{equation}

\subsubsection{Treating $\mathbf{k}_{\mathbf{1}}$ as the control parameter}

\justify
Here, we keep \([A]\), \(k_{2}\), \(k_{3}\), \(k_{4}\), and \([B]\) fixed while varying \(k_{1}\).

\justify
Rearranging equation~\eqref{eq:hopf_condition} gives
\begin{align}
k_{2}\left[ B \right] - k_{4}
&= k_{3}\frac{k_{1}^{2}\left[ A \right]^{2}}{k_{4}^{2}}.
\end{align}
\justify

Thus,
\begin{align}
k_{1}^{2}
&= \frac{k_{4}^{2}}{\left[ A \right]^{2}}
   \frac{k_{2}\left[ B \right] - k_{4}}{k_{3}}.
\end{align}
\justify

Therefore,
\begin{align}
k_{1,H}
&= \frac{k_{4}}{\left[ A \right]}
   \sqrt{\frac{k_{2}\left[ B \right] - k_{4}}{k_{3}}}.
\end{align}
\justify

Again, this requires
\begin{align}
k_{2}\left[ B \right] &> k_{4}.
\end{align}
\justify

The trace is
\begin{align}
\operatorname{tr}\left( J_{0} \right)
&= k_{2}\left[ B \right] - k_{4}
   - k_{3}\frac{k_{1}^{2}\left[ A \right]^{2}}{k_{4}^{2}}.
\end{align}
\justify

Increasing \(k_{1}\) decreases the trace. Therefore,
\begin{align}
k_{1} < k_{1,H}
&\quad \Rightarrow \quad\text{unstable oscillatory state},
\end{align}
\justify
and
\begin{align}
k_{1} > k_{1,H}
&\quad \Rightarrow \quad\text{stable steady state}.
\end{align}
\justify

Thus, for \(k_{1}\) as the control parameter,
\begin{align}
k_{1,H}
&= \frac{k_{4}}{\left[ A \right]}
   \sqrt{\frac{k_{2}\left[ B \right] - k_{4}}{k_{3}}},
\end{align}
\justify
with instability for
\begin{align}
0 < k_{1} &< k_{1,H}.
\end{align}
\justify

\subsubsection{Treating $\mathbf{k}_{\mathbf{2}}$ as the control parameter}

\justify
Here, we keep \([A]\), \([B]\), \(k_{1}\), \(k_{3}\), and \(k_{4}\) fixed while varying \(k_{2}\).

\justify
By rearranging equation~\eqref{eq:hopf_condition}, we get
\begin{align}
k_{2}
&= \frac{1}{\left[ B \right]}
   \left[ k_{4}
   + k_{3}\left( \frac{k_{1}\left[ A \right]}{k_{4}} \right)^{2} \right].
\end{align}
\justify

This assumes
\begin{align}
\left[ B \right] &> 0.
\end{align}
\justify

The trace is
\begin{align}
\operatorname{tr}\left( J_{0} \right)
&= k_{2}\left[ B \right] - k_{4}
   - k_{3}\left( \frac{k_{1}\left[ A \right]}{k_{4}} \right)^{2}.
\end{align}
\justify

Increasing \(k_{2}\) increases the trace. Therefore,
\begin{align}
k_{2} < k_{2,L}
&\quad \Rightarrow \quad\text{stable steady state},
\end{align}
\justify
and
\begin{align}
k_{2} > k_{2,L}
&\quad \Rightarrow \quad\text{oscillatory instability}.
\end{align}
\justify

So for \(k_{2}\) as the control parameter,
\begin{align}
k_{2,L}
&= \frac{k_{4}
+ k_{3}\left( \frac{k_{1}\left[ A \right]}{k_{4}} \right)^{2}}
{\left[ B \right]},
\end{align}
\justify
with instability for
\begin{align}
k_{2} &> k_{2,L}.
\end{align}
\justify

\subsubsection{Treating $\mathbf{k}_{\mathbf{3}}$ as the control parameter}

\justify
Here, we keep \([A]\), \([B]\), \(k_{1}\), \(k_{2}\), and \(k_{4}\) fixed while varying \(k_{3}\).

\justify
By rearranging equation~\eqref{eq:hopf_condition}, we get
\begin{align}
k_{2}\left[ B \right] - k_{4}
&= k_{3}\left( \frac{k_{1}\left[ A \right]}{k_{4}} \right)^{2}.
\end{align}
\justify

Solving for \(k_{3}\),
\begin{align}
k_{3,H}
&= \left( k_{2}\left[ B \right] - k_{4} \right)
   \left( \frac{k_{4}}{k_{1}\left[ A \right]} \right)^{2}.
\end{align}
\justify

This requires
\begin{align}
k_{2}\left[ B \right] &> k_{4}.
\end{align}
\justify

The trace is
\begin{align}
\operatorname{tr}\left( J_{0} \right)
&= k_{2}\left[ B \right] - k_{4}
   - k_{3}\left( \frac{k_{1}\left[ A \right]}{k_{4}} \right)^{2}.
\end{align}
\justify

Increasing \(k_{3}\) decreases the trace. Therefore,
\begin{align}
k_{3} < k_{3,H}
&\quad \Rightarrow \quad\text{oscillatory instability},
\end{align}
\justify
and
\begin{align}
k_{3} > k_{3,H}
&\quad \Rightarrow \quad\text{stable steady state}.
\end{align}
\justify

Thus, for \(k_{3}\) as the control parameter,
\begin{align}
k_{3,H}
&= \left( k_{2}\left[ B \right] - k_{4} \right)
   \left( \frac{k_{4}}{k_{1}\left[ A \right]} \right)^{2},
\end{align}
\justify
with instability for
\begin{align}
0 < k_{3} &< k_{3,H}.
\end{align}
\justify

\subsubsection{Treating $\mathbf{k}_{\mathbf{4}}$ as the control parameter}

\justify
Here, we keep \([A]\), \([B]\), \(k_{1}\), \(k_{2}\), and \(k_{3}\) fixed while varying \(k_{4}\).

\justify
Let
\begin{align}
C &= k_{3}(k_{1}\left[ A \right])^{2},
\end{align}
\justify
and
\begin{align}
\beta &= k_{2}\left[ B \right].
\end{align}
\justify

Then the Hopf condition (Eq. (34)) becomes
\begin{align}
\beta &= k_{4} + \frac{C}{k_{4}^{2}}.
\end{align}
\justify

Multiplying both sides by \(k_{4}^{2}\) gives
\begin{align}
\beta k_{4}^{2} &= k_{4}^{3} + C.
\end{align}
\justify

Therefore,
\begin{align}
k_{4}^{3} - \beta k_{4}^{2} + C &= 0.
\end{align}
\justify

Substituting back into the original parameters,
\begin{align}
k_{4}^{3} - k_{2}\left[ B \right] k_{4}^{2}
+ k_{3}(k_{1}\left[ A \right])^{2}
&= 0.
\end{align}
\justify

Thus, the Hopf values of \(k_{4}\) are the positive roots of this cubic equation.

\justify
the equation is
\begin{align}
k_{4}^{3} - k_{2}\left[ B \right] k_{4}^{2}
+ k_{3}\left( k_{1}\left[ A \right] \right)^{2}
&= 0.
\end{align}
\justify

Define
\begin{align}
a &= k_{2}\left[ B \right],
\quad\quad
c = k_{3}\left( k_{1}\left[ A \right] \right)^{2}.
\end{align}
\justify

Then the equation becomes
\begin{align}
k_{4}^{3} - ak_{4}^{2} + c &= 0.
\end{align}
\justify

Using Cardano's formula, the three formal solutions are
\begin{align}
k_{4,j} ={}&
\frac{a}{3}
+ \omega^{j}
\sqrt[3]{
\frac{a^{3}}{27}
- \frac{c}{2}
+ \sqrt{\frac{c^{2}}{4} - \frac{a^{3}c}{27}}
}
\nonumber\\
&+ \omega^{-j}
\sqrt[3]{
\frac{a^{3}}{27}
- \frac{c}{2}
- \sqrt{\frac{c^{2}}{4} - \frac{a^{3}c}{27}}
}.
\end{align}
\justify

where
\begin{align}
j &= 0,1,2,
\quad\quad
\omega = e^{2\pi i/3}.
\end{align}
\justify

Substituting back,
\begin{align}
k_{4,j} ={}&
\frac{k_{2}\left[ B \right]}{3}
\nonumber\\
&+ \omega^{j}
\sqrt[3]{
\frac{\left( k_{2}\left[ B \right] \right)^{3}}{27}
- \frac{k_{3}\left( k_{1}\left[ A \right] \right)^{2}}{2}
+ \sqrt{
\frac{k_{3}^{2}\left( k_{1}\left[ A \right] \right)^{4}}{4}
- \frac{
\left( k_{2}\left[ B \right] \right)^{3}
k_{3}\left( k_{1}\left[ A \right] \right)^{2}
}{27}
}
}
\nonumber\\
&+ \omega^{-j}
\sqrt[3]{
\frac{\left( k_{2}\left[ B \right] \right)^{3}}{27}
- \frac{k_{3}\left( k_{1}\left[ A \right] \right)^{2}}{2}
- \sqrt{
\frac{k_{3}^{2}\left( k_{1}\left[ A \right] \right)^{4}}{4}
- \frac{
\left( k_{2}\left[ B \right] \right)^{3}
k_{3}\left( k_{1}\left[ A \right] \right)^{2}
}{27}
}
}.
\end{align}
\justify

with
\begin{align}
j &= 0,1,2.
\end{align}
\justify

\justify
For positive real parameters, a useful real-root form is
\begin{align}
k_{4,n} ={}&
\frac{k_{2}\left[ B \right]}{3}
+ \frac{2k_{2}\left[ B \right]}{3}
\cos\left[
\frac{1}{3}
\cos^{-1}\left(
1 - \frac{
27k_{3}\left( k_{1}\left[ A \right] \right)^{2}
}{
2\left( k_{2}\left[ B \right] \right)^{3}
}
\right)
- \frac{2\pi n}{3}
\right].
\end{align}
\justify

where
\begin{align}
n &= 0,1,2.
\end{align}
\justify

This trigonometric form applies when the argument of \(\cos^{-1}\) lies between \(-1\) and \(1\), i.e.
\begin{align}
0 \leq
\frac{
27k_{3}\left( k_{1}\left[ A \right] \right)^{2}
}{
2\left( k_{2}\left[ B \right] \right)^{3}
}
&\leq 2.
\end{align}
\justify

Equivalently,
\begin{align}
0 \leq k_{3}\left( k_{1}\left[ A \right] \right)^{2}
&\leq \frac{4}{27}\left( k_{2}\left[ B \right] \right)^{3}.
\end{align}

\justify
So for \(k_{4}\) as the control parameter,
\begin{align}
k_{4,L} &> k_{4,n}.
\end{align}

\subsubsection{Summary}

Table~\ref{tab:analytic} summarizes the analytical Hopf-bifurcation limits obtained in the previous sections. Substituting the nominal values $[A]=1$, $[B]=3$, and $k_1=k_2=k_3=k_4=1$ into the corresponding stability conditions gives the critical thresholds $[A]=1.414$, $[B]=2$, $k_{1}=1.414$, $k_{2}=0.667$, and $k_{3}=2$. For $k_4$, the Hopf condition yields three real roots, $k_4\approx-0.5321$, $0.6527$, and $2.8794$. Since the negative root is not physically admissible, the two conditions define the oscillatory range $k_4>0.6527$ and $k_4.2.8794$, with the lower value listed in the table.

\begin{table}[H]
\centering
\normalsize
\caption{Conditions for Hopf bifurcation and the corresponding parameter values for $\left[A\right]=1$, $\left[B\right]=3$, and $k_1=k_2=k_3=k_4=1$.}
\label{tab:analytic}
\renewcommand{\arraystretch}{0.5}
\setlength{\tabcolsep}{5pt}
\begin{tabular}{|
>{\centering\arraybackslash}m{0.12\textwidth}|
>{\centering\arraybackslash}m{0.60\textwidth}|
>{\centering\arraybackslash}m{0.10\textwidth}|}
\hline
\textbf{Control Parameter} & \textbf{Condition} & \textbf{Values} \\
\hline

$\left[ A \right]$ &
\[
\left[ A \right]
<
\frac{k_{4}}{k_{1}}
\sqrt{
\frac{k_{2}\left[ B \right]-k_{4}}{k_{3}}
}
\]
&
\[
\left[ A \right] < 1.414
\]
\\
\hline

$\left[ B \right]$ &
\[
\left[ B \right]
>
\frac{1}{k_{2}}
\left[
k_{4}
+
k_{3}
\left(
\frac{k_{1}\left[ A \right]}{k_{4}}
\right)^{2}
\right]
\]
&
\[
\left[ B \right] > 2
\]
\\
\hline

$k_{1}$ &
\[
k_{1}
<
\frac{k_{4}}{\left[ A \right]}
\sqrt{
\frac{k_{2}\left[ B \right]-k_{4}}{k_{3}}
}
\]
&
\[
k_{1} < 1.414
\]
\\
\hline

$k_{2}$ &
\[
k_{2}
>
\frac{
k_{4}
+
k_{3}
\left(
\frac{k_{1}\left[ A \right]}{k_{4}}
\right)^{2}
}{
\left[ B \right]
}
\]
&
\[
k_{2} > 0.667
\]
\\
\hline

$k_{3}$ &
\[
k_{3}
<
\left(
k_{2}\left[ B \right]-k_{4}
\right)
\left(
\frac{k_{4}}{k_{1}\left[ A \right]}
\right)^{2}
\]
&
\[
k_{3} < 2
\]
\\
\hline

$k_{4}$ &
\[
\begin{aligned}
k_{4}
&>
\frac{k_{2}\left[ B \right]}{3}
+
\frac{2k_{2}\left[ B \right]}{3}
\cos\Bigg[
\frac{1}{3}
\cos^{-1}
\left(
1-
\frac{
27k_{3}
\left(
k_{1}\left[ A \right]
\right)^{2}
}{
2
\left(
k_{2}\left[ B \right]
\right)^{3}
}
\right)
-
\frac{2\pi n}{3}
\Bigg]
\end{aligned}
\]
&
\[
k_{4} > 0.6527
\]
\\
\hline

\end{tabular}
\end{table}

\section{Numerical Analysis}
\label{sec:numerical}

The reaction-diffusion equations~\eqref{eq:bruss_pde_X} and~\eqref{eq:bruss_pde_Y} were numerically integrated in a two-dimensional periodic domain using an explicit finite-difference scheme. The concentration fields were initialised with small random perturbations about the homogeneous steady state to seed pattern nucleation; periodic boundary conditions were imposed so that propagating wavefronts re-enter from the opposite edge without artificial reflection. At each time step, the reaction terms were evaluated first, and diffusive coupling to nearest-neighbor grid points was applied thereafter. The two-dimensional field was processed by the following pipeline: frames of 200 $\times$ 200 window size were converted to grayscale and intensity-versus-time signals extracted at four radial sampling points at pixel offsets $x$, $x{+}4$, $x{+}8$, and $x{+}12$ from the centre of an emerging circular pattern.

\begin{figure}
\centering
\includegraphics[width=0.7\textwidth]{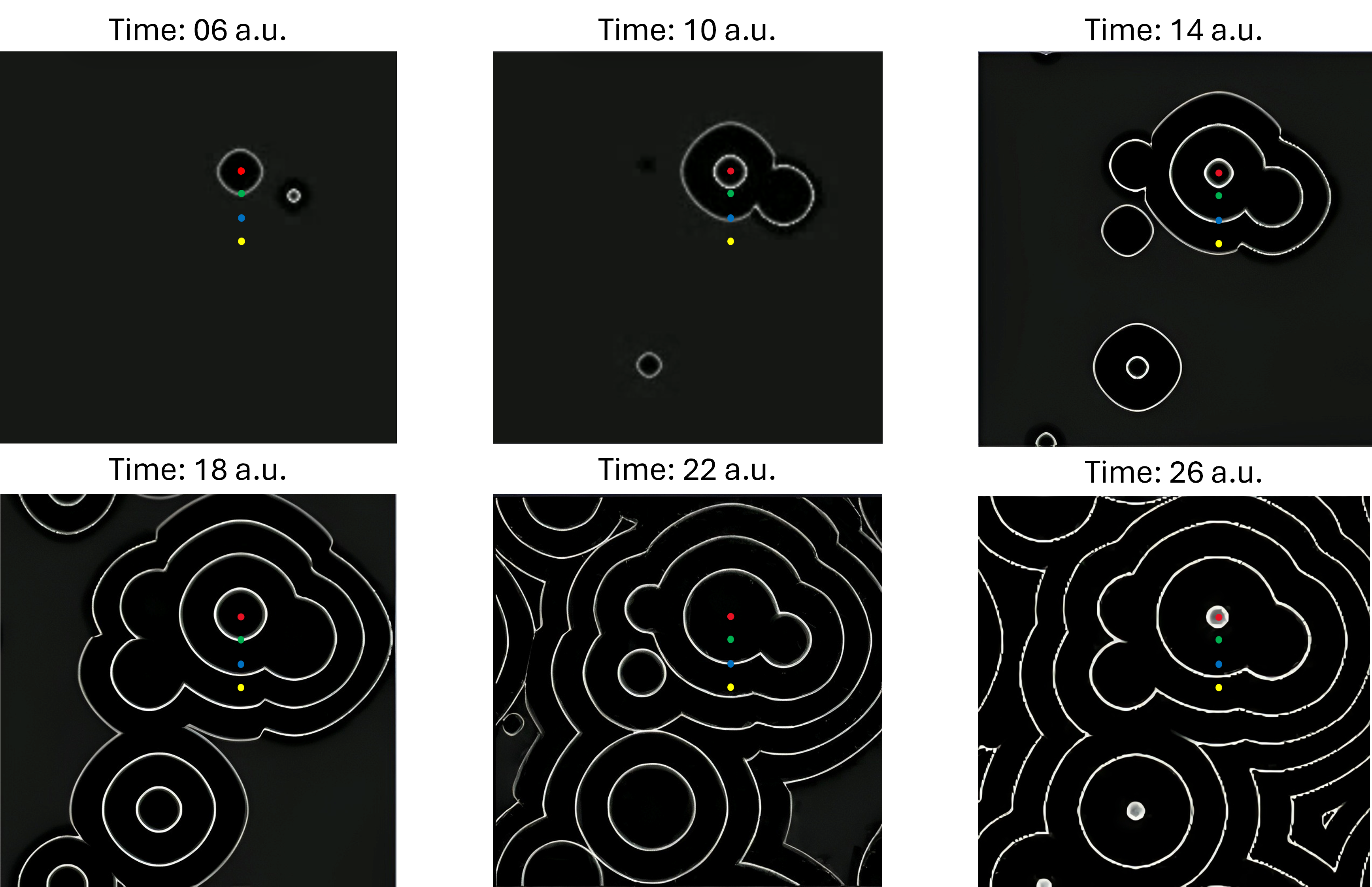}
\caption{Time evolution of the simulated concentration field at $t = 6, 10, 14, 18, 22, 26$ a.u., showing nucleation of a circular wave source and its radial expansion into a system of target patterns, simulated at $[A]=0.6$, $[B]=3$, $k_1=k_2=k_3=k_4=1.0$, $D_X=0.1$, $D_Y=0.01$.}
\label{fig:sim_snapshots}
\end{figure}

The parameter set $[A]=0.6$, $[B]=3$, $k_1=k_2=k_3=k_4=1.0$, $D_X=0.1$, $D_Y=0.01$ places the Brusselator well above the Hopf threshold, in the spatially patterned regime. Under these conditions the concentration field nucleates a circular wave source at early times and develops into expanding concentric target patterns (Figure~\ref{fig:sim_snapshots}), reproducing the spatial morphology characteristic of the BZ reaction. The more informative feature, however, is the temporal waveform at each sampled point: all four locations settle into a stationary train of sharp concentration pulses, with amplitude growing over an initial transient of roughly $10$--$20$~a.u.\ before locking onto a limit cycle. The pronounced anharmonicity of the pulse shape follows directly from the cubic autocatalytic term $k_3[X]^2[Y]$, which drives the oscillation far from sinusoidal once the system sits well above the Hopf threshold; it is this departure from sinusoidal form that generates the harmonic content observed in the frequency domain.

\begin{figure}
\centering
\includegraphics[width=0.7\textwidth]{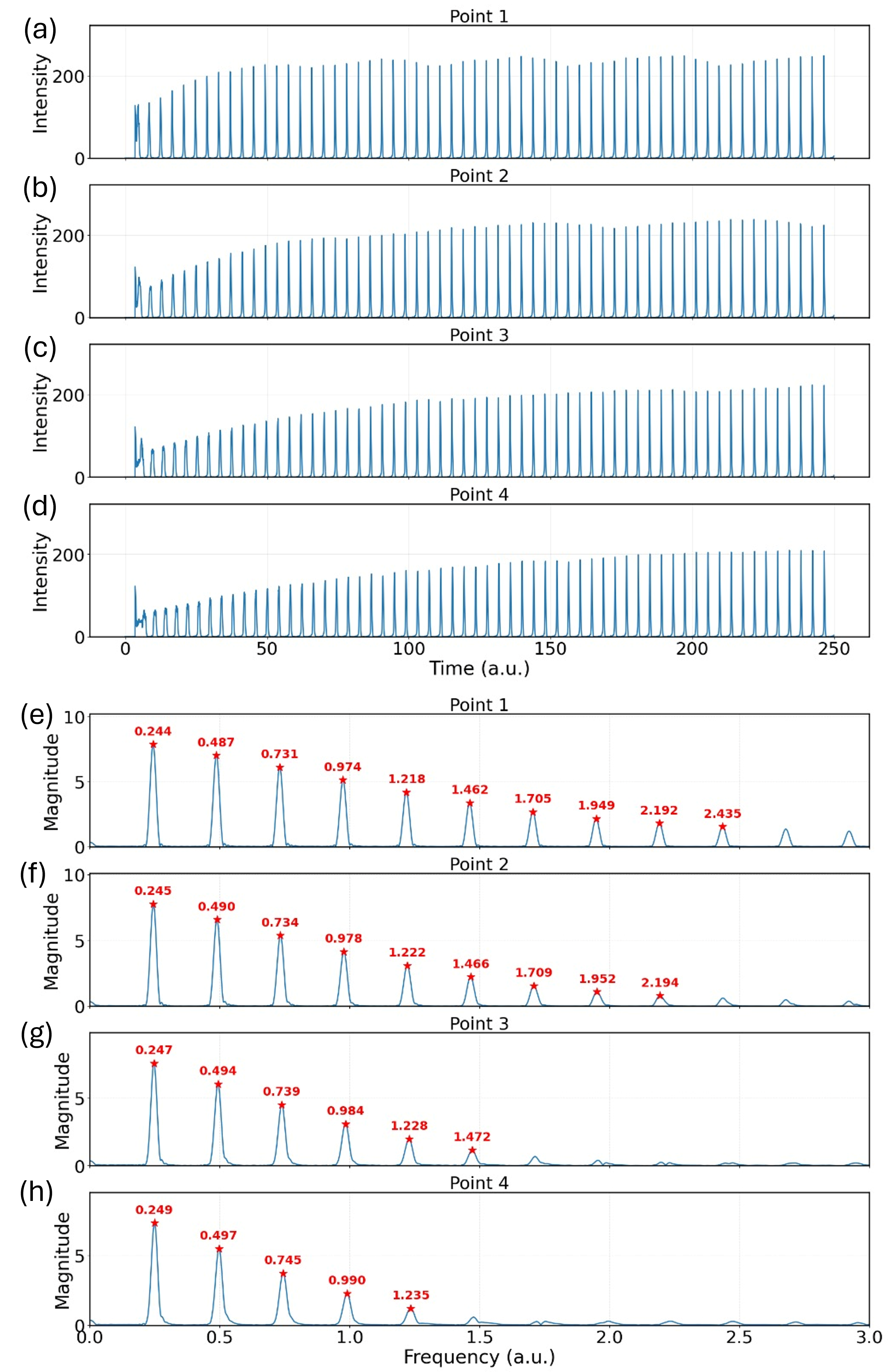}
\caption{Simulated intensity-versus-time traces (a)--(d) and the
corresponding FFT magnitude spectra (e)--(h) at the four radial
sampling points marked in Figure~\ref{fig:sim_snapshots}, obtained
for $[A]=0.6$, $[B]=3$, $k_1=k_2=k_3=k_4=1.0$, $D_X=0.1$, and
$D_Y=0.01$. Both time and frequency are expressed in arbitrary
units.}
\label{fig:sim_fft}
\end{figure}

Each of the four FFT spectra in Figure~\ref{fig:sim_fft}(e)--(h) resolves a dominant fundamental tone in the band $0.244$--$0.249$~a.u., with peak amplitude near $7$--$8$ in normalised intensity units, followed by harmonics at $0.487$, $0.731$, $0.974$, $1.218$, $1.462$, $1.705$, $1.949$, $2.192$, and $2.435$~a.u., extending to at least the tenth harmonic. Spacing between adjacent teeth is uniform to within $2\,\%$ across all resolved harmonics and all four spectra, and harmonic amplitude falls off monotonically with order, from $7$--$8$ at the fundamental to roughly $1$--$2$ at the fifth harmonic and below $0.5$ beyond the eighth --- consistent with progressively weaker energy transfer through successive orders of the trimolecular nonlinearity. This ladder of equally spaced spectral lines, anchored to a well-defined fundamental and generated without any external periodic forcing, is what we refer to throughout this work as the equally spaced spectral grid.

\begin{figure}
\centering
\includegraphics[width=\textwidth]{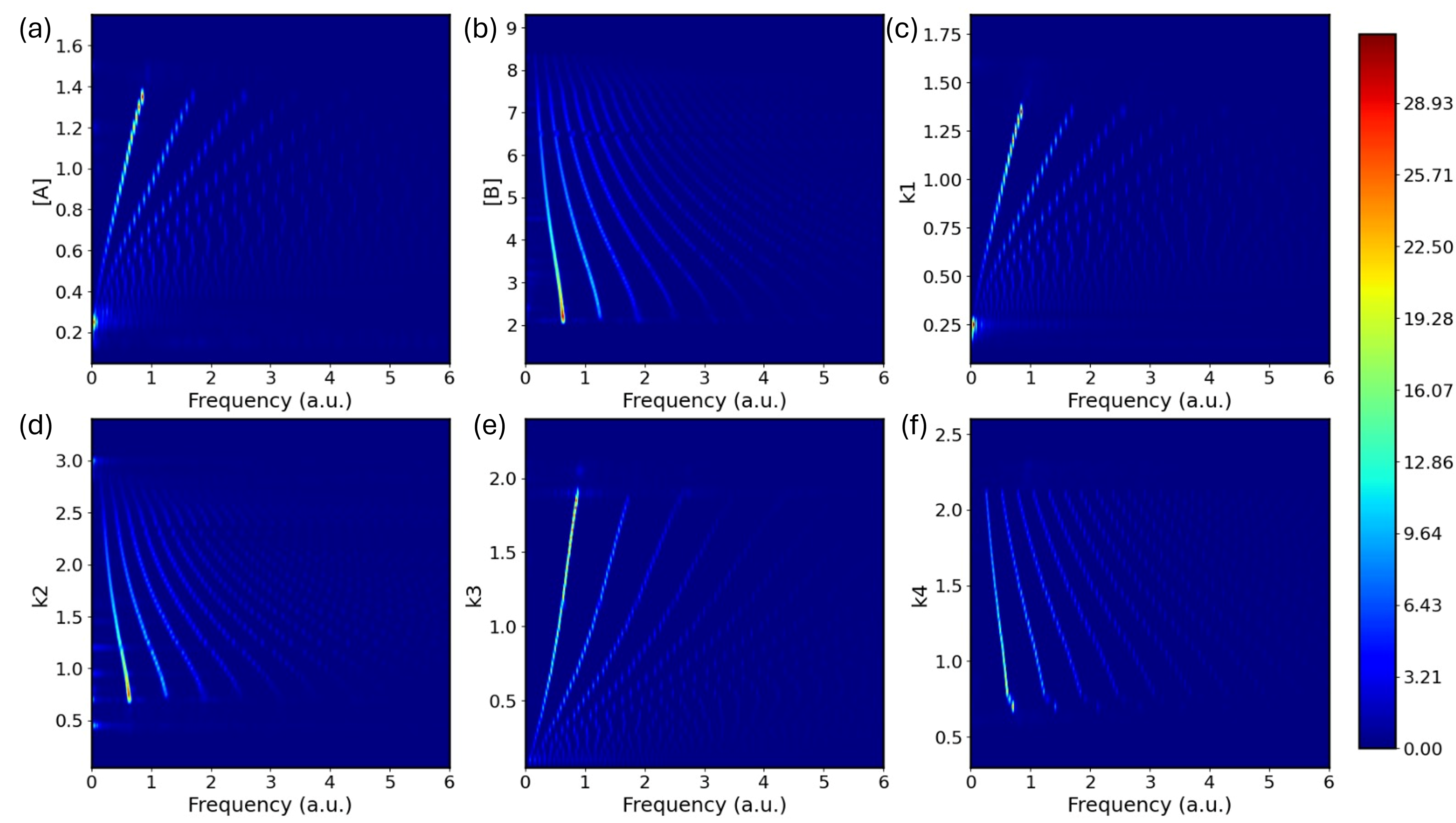}
\caption{Frequency-parameter maps of the simulated FFT amplitude obtained by sweeping (a)~$[A] \in [0.15, 1.4]$, (b)~$[B] \in [2.8, 8.0]$, (c)~$k_1 \in [0.15, 1.6]$, (d)~$k_2 \in [0.45, 3.0]$, (e)~$k_3 \in [0.2, 2.0]$, and (f)~$k_4 \in [0.6, 2.2]$ individually, with all other parameters held at baseline.}
\label{fig:contour}
\end{figure}

To test the robustness of this spectral structure, each of the six parameters $[A]$, $[B]$, $k_1$, $k_2$, $k_3$, and $k_4$ was swept individually over an extended range, with the remaining parameters held at baseline values. The resulting frequency-parameter maps (Figure~\ref{fig:contour}) show that the comb persists across nearly the full sweep in every case; what shifts is only the location of maximum harmonic amplitude. For $[A]$ (Figure~\ref{fig:contour}a), $k_1$ (Figure~\ref{fig:contour}c), and $k_3$ (Figure~\ref{fig:contour}e), the fundamental frequency increases as the parameter increases up to approximately $1.4$, $1.4$ and $1.9$, respectively. For $[B]$ (Figure~\ref{fig:contour}b), $k_2$ (Figure~\ref{fig:contour}d), and $k_4$ (Figure~\ref{fig:contour}f), the fundamental frequency increases as the parameter decreases to approximately $2.1$, $0.7$ and $0.6$, respectively, which generally agree with the values obtained by analytical calculation shown in Table~\ref{tab:analytic}. Across these simulation runs, the inter-harmonic spacing remains uniform throughout, and amplitude and frequency vary smoothly with each parameter.

\begin{figure}
\centering
\includegraphics[width=0.7\textwidth]{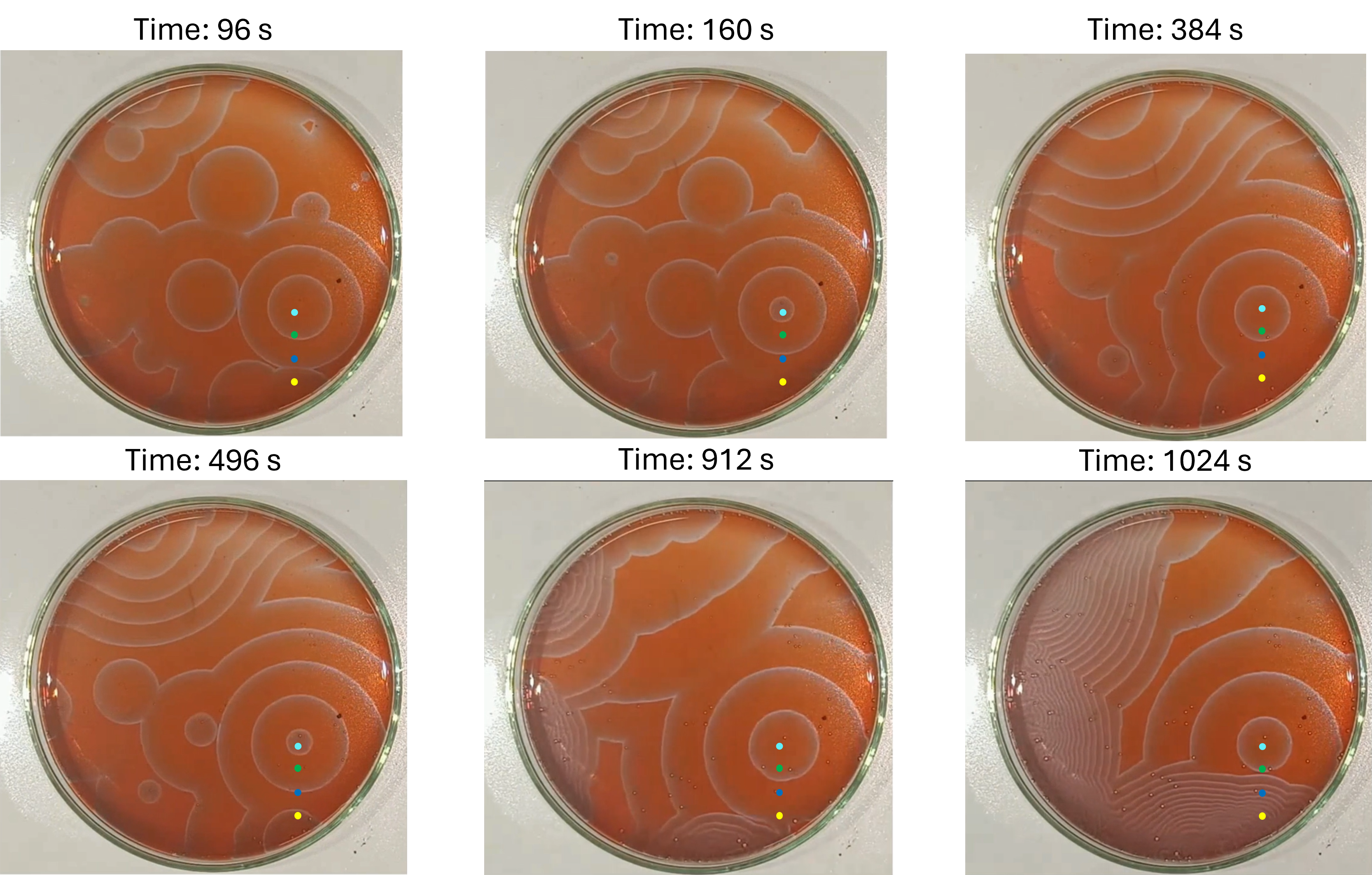}
\caption{BZ reaction in a Petri dish at $t = 96, 160, 384, 496, 912, 1024$~s, showing the formation and propagation of target-wave patterns.}
\label{fig:snapshots}
\end{figure}

\section{Experimental Results}
\label{sec:experiment}

\begin{figure}
\centering
\includegraphics[width=0.7\textwidth]{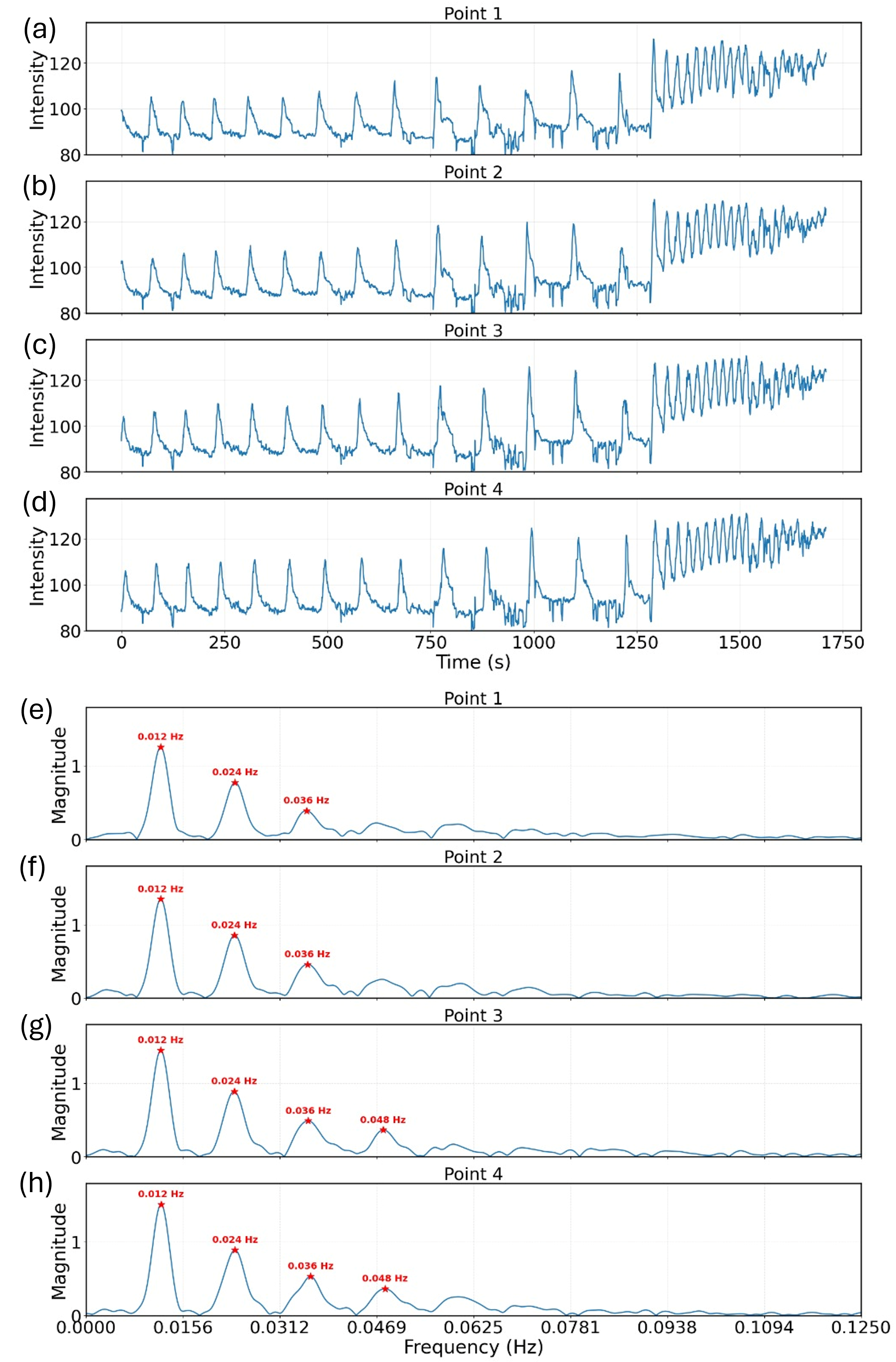}
\caption{Experimental intensity-versus-time traces (a)--(d) and FFT magnitude spectra (e)--(h) at the four radial sampling points marked in Figure~\ref{fig:snapshots}.}
\label{fig:exp_fft}
\end{figure}

The Belousov--Zhabotinsky reaction was run as a thin layer in a Petri dish. Three stock solutions were prepared in advance. Solution A consisted of 65~mL of distilled water, 2~mL of concentrated \ce{H2SO4}, and 5~g of sodium bromate (\ce{NaBrO3}). Solution B was 1~g of malonic acid dissolved in 10~mL of distilled water. Solution C was 1~g of sodium bromide (\ce{NaBr}) in 10~mL of distilled water. These were combined in a 12:2:1 volume ratio, with a small amount of ferroin added so that the redox oscillation appeared as a periodic colour change visible to the camera. The dish was imaged from above using a digital camera under uniform white-LED ring illumination inside a blackout enclosure, which prevented ambient light from interfering with the recorded intensity. Each frame was processed using the same pipeline applied to the simulation output: conversion to grayscale, location of the centre of the circular target pattern as it formed, and extraction of intensity-versus-time traces at four radial points spaced four pixels apart.

Figure~\ref{fig:snapshots} shows the nucleation and propagation of target waves across the dish, confirming that the thin-layer geometry supports the expected reaction-diffusion dynamics throughout the recording. By $t \approx 912$–$1024$~s the background brightness shifts, reflecting a drift in the mean oxidation state of the ferroin indicator as the reagents are depleted; this shift marks the boundary between two dynamically distinct stages in the intensity-time records.

Figures~\ref{fig:exp_fft}(a)--(d) show two successive regimes: a lower-frequency, larger-amplitude regime over the first $\sim\!640$~s, followed by a higher-frequency, smaller-amplitude regime from $t \approx 880$–$960$~s onward. This shift is consistent with progressive consumption of bromate and malonic acid, which shortens the oscillation period as the reaction proceeds. While this phase-transition is interesting, a detailed analysis of this phenomenon is out of the scope of the current manuscript. Hence, we only considered the time series over $0$–$640$~s, and the FFT spectra were obtained for this time span (Figures~\ref{fig:exp_fft}(e)--(h)). Within this window, each of the four spectra shows a dominant peak at $f_0 = 0.012$~Hz with amplitude $1.4$–$1.5$ in normalised units, a second peak at $0.024$~Hz, and a third at $0.036$~Hz; points 3 and 4 also resolve a fourth peak at $0.048$~Hz. The peak spacings are uniform within the frequency resolution of $640$~s, and the harmonic amplitude falls monotonically with order, following the same envelope seen in the simulation.

The measured value, $f_0 = 0.012$~Hz, together with its equally spaced and monotonically decreasing harmonic ladder. The parameter sets associated with the two-variable Brusselator shown in Table~\ref{tab:parameters} reproduce the experimental values with $f_{0,experimental}\sim\frac{1}{16}f_{0,numerical}$. All four sampling points give nearly coincident intensity-time traces over $0$–$640$~s, with no resolvable inter-point delay and no measurable amplitude variation. Spatial averaging or independent oscillations at different radii would produce frequency or phase differences between the four spectra; none are seen. Diffusive coupling phase-locks the oscillation across the sampled region, mirroring the simulation, and the equally spaced peaks are a property of the oscillator.

\begin{table}[H]
\centering
\large
\caption{Parameter sets used in the simulations.}
\label{tab:parameters}
\renewcommand{\arraystretch}{1.25}

\begin{tabular}{|c|c|c|c|c|c|c|c|}
\hline
\textbf{$[A]$} &
\textbf{$[B]$} &
\textbf{$k_1$} &
\textbf{$k_2$} &
\textbf{$k_3$} &
\textbf{$k_4$} &
\textbf{$f_{0,numerical}$ (a.u.)} &
\textbf{$f_{0,experimental}$ (Hz)} \\
\hline

0.53 & 3   & 1     & 1    & 1     & 1    & 0.192 & 0.012 \\
\hline
1    & 7.8 & 1     & 1    & 1     & 1    & 0.192 & 0.012 \\
\hline
1    & 3   & 0.525 & 1    & 1     & 1    & 0.192 & 0.012 \\
\hline
1    & 3   & 1     & 2.58 & 1     & 1    & 0.192 & 0.012 \\
\hline
1    & 3   & 1     & 1    & 0.275 & 1    & 0.192 & 0.012 \\
\hline
1    & 3   & 1     & 1    & 1     & 2.54 & 0.192 & 0.012 \\
\hline

\end{tabular}
\end{table}

\section{Conclusion}
\label{sec:conclusions}

The Brusselator model, set above its Hopf threshold, produces a frequency comb. The simulated spectrum shows a fundamental frequency together with an evenly spaced harmonic ladder resolved out to high order, its amplitude falling off monotonically as the order increases. This behaviour comes entirely from the trimolecular term $k_3[X]^2[Y]$: as the sole nonlinearity in the kinetics, it both sets the fundamental frequency and opens the intermodal mixing channel that fills in the ladder. Single-parameter sweeps over the rate constants and feed concentrations confirm that the ladder is a structural feature rather than something fine-tuned into existence --- it survives across every parameter range tested.

The Belousov--Zhabotinsky reaction reproduces this behavior in a real chemical system. The measured spectrum shows a fundamental frequency together with three to four equally spaced, monotonically decaying harmonics, matching the Brusselator model output for every parameter set tested, with no chemistry-specific adjustment. The four radial sampling points give effectively coincident spectra, showing that diffusive coupling phase-locks the oscillation across the imaged region. Together, these results extend the frequency-comb concept to chemical kinetics. The same ingredients seen in other physical settings, namely nonlinear mode coupling, self-consistent energy redistribution, and phase locking, are at work here as well, with the cubic autocatalytic term of the Brusselator playing the role filled elsewhere by mechanical, magnetic, or electromagnetic nonlinearities. Reaction-diffusion oscillators thus stand as the chemical counterpart in this broader class of systems governed by the same organizing principle on a different physical realm.

Two extensions follow from these findings. A higher-resolution event camera at a fixed observation point should resolve more comb fingers. A systematic parametric study, with temperature included through its Arrhenius effect on the rate constants, would map how the comb spacing, harmonic amplitudes, spectral bandwidth, and onset threshold depend on the operating point. Chemical frequency combs are also of practical interest in their own right: drift in reactor conditions appears as a measurable shift in the comb spacing and harmonic content, offering a route to real-time monitoring and control. The same sensitivity to kinetics and transport supports chemical and environmental sensing, including contaminant detection and tracking of compositional change in complex mixtures.

\printbibliography

@book{Nicolis1977,
  author    = {Gregoire Nicolis and Ilya Prigogine},
  title     = {Self-Organization in Nonequilibrium Systems: From Dissipative Structures to Order through Fluctuations},
  publisher = {Wiley-Interscience},
  address   = {New York},
  year      = {1977}
}

@article{Turing1952,
  author  = {Alan M. Turing},
  title   = {The chemical basis of morphogenesis},
  journal = {Philosophical Transactions of the Royal Society of London B},
  volume  = {237},
  pages   = {37--72},
  year    = {1952},
  doi     = {10.1098/rstb.1952.0012}
}

@misc{Belousov1959,
  author    = {Boris P. Belousov},
  title     = {A periodic reaction and its mechanism},
  note      = {Collection of Abstracts on Radiation Medicine, pp. 145--147},
  year      = {1959}
}

@inproceedings{zhabotinsky1964periodic,
  title={Periodic oxidizing reactions in the liquid phase},
  author={Zhabotinsky, Anatolii Markovich},
  booktitle={Doklady Akademii Nauk},
  volume={157},
  number={2},
  pages={392--395},
  year={1964},
  organization={Russian Academy of Sciences}
}

@article{ZaikinZhabotinsky1970,
  author    = {Zaikin, A. N. and Zhabotinsky, A. M.},
  title     = {Concentration Wave Propagation in Two-Dimensional
               Liquid-Phase Self-Oscillating System},
  journal   = {Nature},
  year      = {1970},
  volume    = {225},
  number    = {5232},
  pages     = {535--537},
  doi       = {10.1038/225535b0}
}

@article{Field1972,
  author    = {Field, Richard J. and K{\H{o}}r{\"{o}}s, Endre and Noyes, Richard M.},
  title     = {Oscillations in Chemical Systems. {II}.
               Thorough Analysis of Temporal Oscillation in the
               Bromate--Cerium--Malonic Acid System},
  journal   = {Journal of the American Chemical Society},
  year      = {1972},
  volume    = {94},
  number    = {25},
  pages     = {8649--8664},
  doi       = {10.1021/ja00780a001}
}

@article{KondoMiura2010,
  author    = {Kondo, Shigeru and Miura, Takashi},
  title     = {Reaction-Diffusion Model as a Framework for Understanding
               Biological Pattern Formation},
  journal   = {Science},
  year      = {2010},
  volume    = {329},
  number    = {5999},
  pages     = {1616--1620},
  doi       = {10.1126/science.1179047}
}

@article{HalatekFrey2018,
  author    = {Halatek, Jacob and Frey, Erwin},
  title     = {Rethinking Pattern Formation in Reaction--Diffusion Systems},
  journal   = {Nature Physics},
  year      = {2018},
  volume    = {14},
  number    = {5},
  pages     = {507--514},
  doi       = {10.1038/s41567-017-0040-5}
}

@article{BraunsFrey2020,
  author    = {Brauns, Fridtjof and Halatek, Jacob and Frey, Erwin},
  title     = {Phase-Space Geometry of Mass-Conserving
               Reaction-Diffusion Dynamics},
  journal   = {Physical Review X},
  year      = {2020},
  volume    = {10},
  number    = {4},
  pages     = {041036},
  doi       = {10.1103/PhysRevX.10.041036}
}

@article{MarciniakFrey2016,
  author    = {Marciniak-Czochra, Anna and Karch, Grzegorz and Suzuki, Kanako},
  title     = {Instability of {Turing} Patterns in Reaction-Diffusion-{ODE} Systems},
  journal   = {Journal of Mathematical Biology},
  year      = {2017},
  volume    = {74},
  number    = {3},
  pages     = {583--618},
  doi       = {10.1007/s00285-016-1035-z}
}

@article{CupicEpstein2021,
  author    = {Cupi{\'c}, {\v{Z}}eljko D. and Taylor, Annette F.
               and Horv{\'a}th, Dezso and Orlik, Marek and Epstein, Irving R.},
  title     = {Editorial: Advances in Oscillating Reactions},
  journal   = {Frontiers in Chemistry},
  year      = {2021},
  volume    = {9},
  pages     = {690699},
  doi       = {10.3389/fchem.2021.690699}
}

@article{DalwadiPearce2023,
  author    = {Dalwadi, Mohit P. and Pearce, Philip},
  title     = {Universal Dynamics of Biological Pattern Formation in
               Spatio-Temporal Morphogen Variations},
  journal   = {Proceedings of the Royal Society A:
               Mathematical, Physical and Engineering Sciences},
  year      = {2023},
  volume    = {479},
  number    = {2271},
  pages     = {20220829},
  doi       = {10.1098/rspa.2022.0829}
}

@misc{WuZhang2024,
  title={Solution landscape of reaction-diffusion systems reveals a nonlinear mechanism and spatial robustness of pattern formation},
  author={Wu, Shuonan and Yu, Bing and Tu, Yuhai and Zhang, Lei},
  journal={Fundamental Research},
  year={2025},
  publisher={Elsevier}
}

@article{FieldNoyes1974,
  author  = {Richard J. Field and Richard M. Noyes},
  title   = {Oscillations in chemical systems. {IV}. {L}imit cycle behavior in a model of a real chemical reaction},
  journal = {Journal of Chemical Physics},
  volume  = {60},
  pages   = {1877--1884},
  year    = {1974},
  doi     = {10.1063/1.1681288}
}

@book{Epstein1998,
  author    = {Irving R. Epstein and John A. Pojman},
  title     = {An Introduction to Nonlinear Chemical Dynamics: Oscillations, Waves, Patterns, and Chaos},
  publisher = {Oxford University Press},
  address   = {New York},
  year      = {1998}
}

@book{Strogatz2014,
  author    = {Steven H. Strogatz},
  title     = {Nonlinear Dynamics and Chaos: With Applications to Physics, Biology, Chemistry, and Engineering},
  edition   = {2nd},
  publisher = {Westview Press},
  address   = {Boulder, CO},
  year      = {2014}
}

@article{Prigogine1968,
  author  = {Ilya Prigogine and René Lefever},
  title   = {Symmetry breaking instabilities in dissipative systems. {II}},
  journal = {Journal of Chemical Physics},
  volume  = {48},
  pages   = {1695--1700},
  year    = {1968},
  doi     = {10.1063/1.1668896}
}

@article{Udem2002,
  author  = {Thomas Udem and Ronald Holzwarth and Theodor W. Hänsch},
  title   = {Optical frequency metrology},
  journal = {Nature},
  volume  = {416},
  pages   = {233--237},
  year    = {2002},
  doi     = {10.1038/416233a}
}

@article{Cundiff2003,
  author  = {Steven T. Cundiff and Jun Ye},
  title   = {Colloquium: {F}emtosecond optical frequency combs},
  journal = {Reviews of Modern Physics},
  volume  = {75},
  pages   = {325--342},
  year    = {2003},
  doi     = {10.1103/RevModPhys.75.325}
}

@article{Ganesan2017,
  author  = {Adarsh Ganesan and Cuong Do and Ashwin Seshia},
  title   = {Phononic frequency comb via intrinsic three-wave mixing},
  journal = {Physical Review Letters},
  volume  = {118},
  pages   = {033903},
  year    = {2017},
  doi     = {10.1103/PhysRevLett.118.033903}
}

@article{LengyelEpstein1991,
  author  = {Lengyel, Istv{\'a}n and Epstein, Irving R.},
  title   = {Modeling of {Turing} Structures in the Chlorite--Iodide--Malonic Acid--Starch Reaction System},
  journal = {Science},
  volume  = {251},
  number  = {4994},
  pages   = {650--652},
  year    = {1991},
  doi     = {10.1126/science.251.4994.650}
}

@article{EpsteinShowalter1996,
  author  = {Epstein, Irving R. and Showalter, Kenneth},
  title   = {Nonlinear Chemical Dynamics: Oscillations, Patterns, and Chaos},
  journal = {The Journal of Physical Chemistry},
  volume  = {100},
  number  = {31},
  pages   = {13132--13147},
  year    = {1996},
  doi     = {10.1021/jp953547m}
}

@article{DeKepperEpstein1981,
  author  = {De Kepper, Patrick and Epstein, Irving R. and Kustin, Kenneth},
  title   = {A Systematically Designed Homogeneous Oscillating Reaction: The Arsenite--Iodate--Chlorite System},
  journal = {Journal of the American Chemical Society},
  volume  = {103},
  number  = {8},
  pages   = {2133--2134},
  year    = {1981},
  doi     = {10.1021/ja00398a061}
}

@article{VanagEpstein2009,
  author  = {Vanag, Vladimir K. and Epstein, Irving R.},
  title   = {Cross-Diffusion and Pattern Formation in Reaction--Diffusion Systems},
  journal = {Physical Chemistry Chemical Physics},
  volume  = {11},
  number  = {6},
  pages   = {897--912},
  year    = {2009},
  doi     = {10.1039/B813825G}
}

@article{Epstein2014,
  author  = {Epstein, Irving R.},
  title   = {Coupled Chemical Oscillators and Emergent System Properties},
  journal = {Chemical Communications},
  volume  = {50},
  number  = {74},
  pages   = {10758--10767},
  year    = {2014},
  doi     = {10.1039/C4CC00290C}
}

@article{AwalEpstein2023,
  author  = {Awal, Naziru M. and Epstein, Irving R. and Kaper, Tasso J. and Vo, Theodore},
  title   = {Symmetry-Breaking Rhythms in Coupled, Identical Fast--Slow Oscillators},
  journal = {Chaos: An Interdisciplinary Journal of Nonlinear Science},
  volume  = {33},
  number  = {1},
  pages   = {011102},
  year    = {2023},
  doi     = {10.1063/5.0131305}
}

@article{EpsteinAwal2024,
  author  = {Epstein, Irving R. and Awal, Naziru M. and Kaper, Tasso J. and Vo, Theodore},
  title   = {Strong Symmetry Breaking Rhythms Created by Folded Nodes in a Pair of Symmetrically Coupled, Identical {Koper} Oscillators},
  journal = {Chaos: An Interdisciplinary Journal of Nonlinear Science},
  volume  = {34},
  number  = {5},
  pages   = {053142},
  year    = {2024},
  doi     = {10.1063/5.0202872}
}

@article{Ganesan2018a,
  author  = {Adarsh Ganesan and Cuong Do and Ashwin Seshia},
  title   = {Phononic frequency comb via three-mode parametric resonance},
  journal = {Applied Physics Letters},
  volume  = {112},
  pages   = {021906},
  year    = {2018},
  doi     = {10.1063/1.5003133}
}

@article{Ganesan2018b,
  author  = {Adarsh Ganesan and Cuong Do and Ashwin Seshia},
  title   = {Excitation of coupled phononic frequency combs via two-mode parametric three-wave mixing},
  journal = {Physical Review B},
  volume  = {97},
  pages   = {014302},
  year    = {2018},
  doi     = {10.1103/PhysRevB.97.014302}
}

@article{deJong2023,
  author  = {Matthijs H. J. de Jong and Adarsh Ganesan and Andrea Cupertino and Simon Gröblacher and Richard A. Norte},
  title   = {Mechanical overtone frequency combs},
  journal = {Nature Communications},
  volume  = {14},
  pages   = {1458},
  year    = {2023},
  doi     = {10.1038/s41467-023-36953-8}
}

@article{Qi2020,
  author  = {Zhen Qi and Curtis R. Menyuk and Jason J. Gorman and Adarsh Ganesan},
  title   = {Existence conditions for phononic frequency combs},
  journal = {Applied Physics Letters},
  volume  = {117},
  pages   = {183501},
  year    = {2020},
  doi     = {10.1063/5.0025314}
}

@article{Kesekler2022,
  author  = {Ata Kesekler and Hadi Arjmandi-Tash and Peter G. Steeneken and Farbod Alijani},
  title   = {Symmetry-breaking-induced frequency combs in graphene resonators},
  journal = {Nano Letters},
  volume  = {22},
  number  = {15},
  pages   = {6048--6054},
  year    = {2022},
  publisher = {ACS Publications},
  doi     = {10.1021/acs.nanolett.2c02074}
}

@article{Chiout2021,
  author  = {Anis Chiout and others},
  title   = {Multi-order phononic frequency comb generation within a {MoS$_2$} electromechanical resonator},
  journal = {Applied Physics Letters},
  volume  = {119},
  pages   = {173102},
  year    = {2021}
}

@article{anderson2026phononic,
  author  = {Ian Anderson and Jack Kramer and Tzu-Hsuan Hsu and Yinan Wang and Vakhtang Chulukhadze and Ruochen Lu},
  title   = {Phononic combs in lithium niobate acoustic resonators},
  journal = {Applied Physics Letters},
  volume  = {128},
  number  = {5},
  year    = {2026},
  publisher = {AIP Publishing}
}

@article{Liu2025,
  author  = {Rumeng Liu and Guangfei Zhu},
  title   = {Phononic frequency combs in twisted bilayer van der {W}aals material resonators},
  journal = {Journal of Applied Physics},
  volume  = {138},
  pages   = {144305},
  year    = {2025},
  doi     = {10.1063/5.0296669}
}

@article{Lei2024,
  author  = {Hongbin Lei and Qian Zhang and Hongqiang Xie and Congsen Meng and Zhaoyang Peng and Xiaowei Wang and Jinlei Liu and Guangru Bai and Adarsh Ganesan and Zengxiu Zhao},
  journal = {Optics Express},
  number  = {3},
  pages   = {5396--5410},
  publisher = {Optica Publishing Group},
  title   = {On the nonlinear rovibrational excitation of phononic frequency combs in molecules},
  volume  = {33},
  month   = {Feb},
  year    = {2025},
  doi     = {10.1364/OE.544325}
}

@article{Rangwala2026,
  author  = {Murtaza Rangwala and Adarsh Ganesan},
  title   = {Spontaneous symmetry breaking and collective {H}iggs--{G}oldstone dynamics in solid-state phononic frequency combs},
  journal = {arXiv preprint arXiv:2602.07462},
  year    = {2026}
}

@article{Xiao2026,
  author  = {Guangzong Xiao and Zijian Feng and Tengfang Kuang and Ran Huang and Yutong He and Xinlin Chen and Yunlan Zuo and Xiang Han and Wei Xiong and Zhongqi Tan and Adarsh Ganesan and Franco Nori and Cheng-Wei Qiu and Xiaobao Zhang and Hui Luo and Hui Jing},
  title   = {Ultrabroadband phonon laser frequency comb},
  journal = {Advanced Photonics},
  volume  = {8},
  number  = {2},
  pages   = {026004},
  year    = {2026},
  doi     = {10.1117/1.AP.8.2.026004}
}

@article{Demidov2020,
  author  = {Vasyl E. Demidov and Vladislav I. Urazhdin and Sergei O. Demokritov},
  title   = {Sustained coherent spin wave emission using frequency combs},
  journal = {Physical Review B},
  volume  = {101},
  pages   = {224423},
  year    = {2020},
  doi     = {10.1103/PhysRevB.101.224423}
}

@article{PolarizationComb2026,
  author  = {Xiyin Ye and Tao Yu},
  title   = {Frequency comb of electric-polarization waves},
  journal = {arXiv preprint arXiv:2603.27947},
  year    = {2026}
}

@article{Trivedi2026,
  author  = {Oem Trivedi and Madhurendra Mishra and Adarsh Ganesan},
  title   = {Cosmological frequency combs},
  journal = {Europhysics Letters},
  year    = {2026},
  doi     = {10.1209/0295-5075/ae55c4}
}
\end{document}